# Simulation of Non-linear SRF Losses Derived from Characteristic Topography of Etched and Electropolished Niobium Surfaces

Chen Xu[a,b], Charles E. Reece[a] and Michael J. Kelley[a, b]

[a] Thomas Jefferson National Accelerator Facility, Newport News, VA 23606, USA

[b] Dept. of Applied Science. The College of William & Mary, Williamsburg, VA 23185, USA

**Abstract:**

A simplified numerical model has been developed to simulate non-linear superconducting radiofrequency (SRF) losses on Nb surfaces. This study focuses exclusively on excessive surface resistance ($R_s$) losses due to the microscopic topographical magnetic field enhancements. When the enhanced local surface magnetic field exceeds the superconducting critical transition magnetic field $H_c$, small volumes of surface material may become normal conducting and increase the effective surface resistance without inducing a quench. We seek to build an improved quantitative characterization of this qualitative model. Using topographic data from typical Buffered Chemical Polish (BCP) and Electropolish (EP) treated fine grain niobium, we have estimated the resulting field-dependent losses and extrapolated this model to the implications for cavity performance. The model predictions correspond well to the characteristic BCP versus EP high field $Q_0$ performance differences for fine grain niobium. We describe the algorithm of the model, its limitations, and the effects of this non-linear loss contribution on SRF cavity performance.

1. Introduction

The roughness of SRF surfaces has long been recognized as influential on mechanisms which limit the performance of niobium SRF resonators. [1] Decreasing cavity unloaded quality factor $Q_0$ reflects an increasing of average surface resistance, $R_s$. Several models attempt to explain the dependency of quality factors at different accelerator fields. Agreement of these models and experiments has been mixed. [2] Surface roughness has been associated with increased losses, lower quench fields, and increased difficulty in cleaning. Various surface treatments are implemented to achieve beneficial smoothness. Typical surfaces have been statistically characterized and analyzed. Different treatments modify surface features at various lateral length scales. [3,4] However, it has not been well established just how the details of topographical features directly affect integrated RF performance. It is understood that sharp features promote magnetic field enhancement and may, under appropriate conditions, initiate quench. [5,6].

Knobloch et al. estimated RF loss from a grain boundary edge and extended this estimate to anticipated effects in an SRF cavity. [5] Here, this method is improved by a detailed finite element method simulation. This simulation also integrates both RF field and thermal calculations on representative niobium SRF surfaces obtained by AFM profilometry with micrometer resolution rather than infer a distribution function of local field enhancements from observed cavity performance constraints. In addition, we allow the size of local normal conducting volumes to be determined dynamically, rather than assuming a fixed width and depth as was done in [5]. Secondly, this simulation incorporates the temperature dependency of various superconducting material properties. We customized an algorithm to iteratively compute RF losses under steady state conditions. Our analysis provides for no fit parameters, just direct calculation limited by the available mesh resolution. Such an attempt to model increased RF losses due to topographic enhancements has not been previously reported. Thirdly, we will relate the simulation results to accelerator cavity performance differences associated with either chemical etching or electropolishing finishing steps.

Typical BCP-treated fine-grain Nb cavities commonly show a $Q_0$ that starts to decrease with dramatic slope when the accelerating gradient increases from 16 MV/m to 22 MV/m. This occurs even after the cavities are treated with a post-chemistry bake. [7–9] In some extreme cases, cavities exhibit this kind of nonlinear loss when the accelerating gradient is as low as 15 MV/m. After EP treatment, this $Q_0$ decrease is removed. [9-11] This frequently encountered phenomenon is dramatically illustrated in Figure 1, which presents the performance of Jefferson Lab CEBAF 7-cell prototype cavity HG006 with very heavy BCP etch followed by a 30 µm EP, with no field emission loading in any test. [12] Such a difference in performance has come to be qualitatively associated with field enhancements of the "rougher" BCP-treated surface. Such roughness from chemical etching can be highly variable depending on crystalline structure and defect density of the niobium surface and amount of material removal. Since the principal difference between these two surface states is microscopic topographical roughness, these results suggest that managing topography evolution plays a critical role in improving useful cavity gradient.

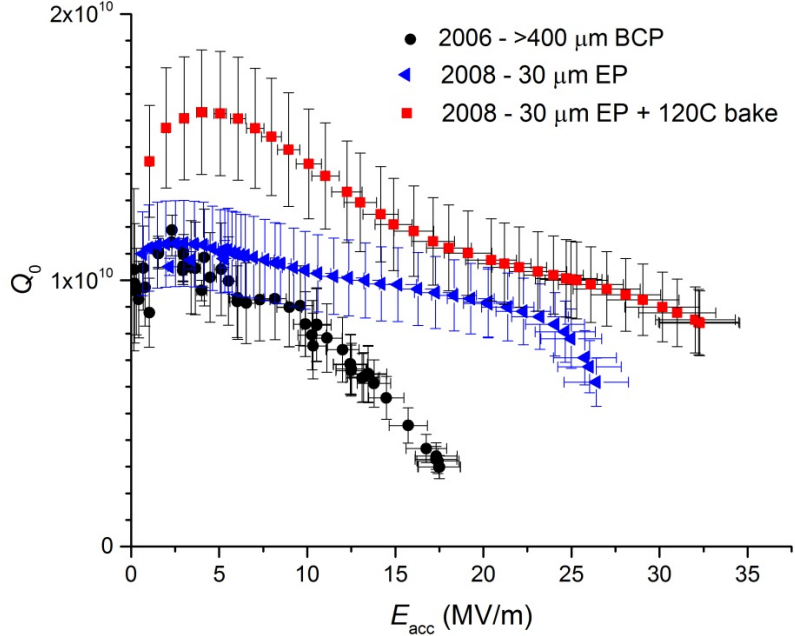

Figure 1: Performance of Nb cavity HG006 after a very heavy etching BCP and a subsequent 30 μm EP [12].

Compared to EP surfaces, BCP treated surfaces have a higher root mean square (RMS) height variation and a greater density of sharp features.[3,4] Those high and sharp features enhance the local magnetic field, and the enhanced magnetic field at these local features may exceed the superconducting critical field $H_{crit}$. As a result, local superconducting transition is initiated and small areas may become normal conducting. For niobium, in general a mixed state occurs and transition is quite complicated. As a weak type II superconductor, Nb has a Ginzburg-Landau factor around 1.3, close to type I superconductors. [13]

In this paper, we focus our study on the topographical enhancements to surface RF losses. For simplification, we simply treat Nb as a type I superconductor. Thus, a mixed state will be ignored and Nb will become normal conducting whenever the local field exceeds $H_c$, as a type I superconductor. We choose to simplify the normal to superconducting transition by using a single value $H_{crit}$ to identify normal and superconducting boundary. In a more complex case, a section of material under the surface will be in a mixed state when $H_{crit}$ is reached. Thus the normal and superconducting boundary will become a belt rather than a line. This belt volume represents the mixed state. With added complexity one could calculate the RF loss within this belt and the normal-conducting zones. We leave that effort for future research. In this analysis, the precise value of $H_{crit}$ is open for discussion; somewhat arbitrarily we use the superheating field $H_{sh}$ in our simulations. [14] Note that this $H_{sh}$ changes dynamically according to the local temperature and also may be suppressed via the mechanism described by Kubo [15]. Uncertainty in the precise value of the effective $H_{crit}$ has little impact on the resulting analysis of the present work, but remains an opportunity for further future refinement.

The local effective field enhancement may be quantified by the Local Geometric Magnetic Field Enhancement factor (LGMFE). This index is a ratio of local enhanced magnetic field over the nominally applied RF field. [16,17]

The magnetic field amplitude decays exponentially in the Nb material. When the surface $H$ field is greater than $H_c$, a location inside the surface will have an $H$ field less than $H_c$. In this circumstance, there is an interface between normal conducting material on the surface and superconducting material in the bulk. Because the electric time constants are so short compared with the RF frequency, this interface is moving along with the RF phase in our simulation relevant to 1-2 GHz cavities. An excess RF loss is generated by these small normal nucleation sites on the surface. Moreover, this RF loss raises the local surface temperature and consequently reduces the local $H_c$. The positive thermal feedback aggravates the normal conducting transition. Detailed calculation is needed to evaluate the local RF loss and attendant temperature rise. The consequential additional RF loss can be expressed as an increasing effective surface resistance.[18,19] In addition, a temperature map may be calculated to estimate the local $H_c$. Temperature rise would increase the normal zones and bring additional loss. In this analysis, electromagnetic and thermal iterations are adapted to mimic this thermal feedback condition. Stable solutions are approached with a convergence. We propose a model to calculate non-linear RF loss from microscopic surface topographical features. An averaged surface resistance as a function of applied *H* field is given to compare with cavity cold testing experiments.

In this analysis, electromagnetic and thermal simulations are numerically provided by the Finite Element Method (FEM). Corresponding field-dependent RF ohmic losses are characterized from surface topography associated with two types of popular surface treatments. The effective $R_s$ values are calculated, and corresponding quality factor, *Q*, versus accelerating gradient, *E*, curves are generated from this analytic model. The model may be applied to cavities with various surface treatments in order to further understand and predict the influence of surface topography on practical resonators at high surface magnetic fields.

## 2. Methodology

### 2.1 Electromagnetic calculation

**Electromagnetism FEM:**

To calculate the electromagnetic field distribution near a surface, Maxwell's equations must be solved with a boundary condition by an eigenmode solver.[20] We reorganize the Maxwell equations into a Helmholtz equation as shown:

$$\left(\nabla^2 + k^2\right)\phi = 0 \qquad (1)$$

where $\varphi$ is the magnetic scalar potential and wavenumber $k = |k| = \frac{\omega}{c}$.

After separation of variables, space $\phi(r)$ and time $T(t)$ give general wave solutions. An example of 1D solution is expressed below:

$$\Phi(r) = \sum_n C_1 e^{ik_n \cdot r} + \sum_n C_2 e^{-ik_n \cdot r}$$
$$T(t) = \sum_n D_1 e^{i\omega_n \cdot t} + \sum_n D_2 e^{-i\omega_n \cdot t} \quad \text{............................} \quad (2)$$

In our case, we simplify the wave equation into a static form near the surface. The simplification is appropriate when the second term in equation 1 is much smaller than the first term. This is applicable when the simulated area lateral size is much smaller than the RF wavelength. In our simulation, the lateral scale $l$ is 100 µm while wavelength $\lambda$ at 1.5 GHz is 20 cm. At this simulated scale, the wavenumber $k$ has an order $10^{-2}$ cm$^{-1}$. The sinusoidal field difference within the simulated length is trivial. Therefore, the dominating equation reduces into a Laplace equation, given in equation 3.

$$\nabla^2 \varphi = 0 \quad \text{.....................................} \quad (3)$$

FEM and conformal mapping methods are used to solve the Laplace equation in 2D. The RF $H$ field gets enhanced when it crosses the groove features and remains uniform when the $H$ field comes along the groove direction. Note that $\varphi$ in equation 2 and 3 can be interchanged with any vector fields and scalar potentials, such as electric field $E$, magnetic field $H$, magnetic flux $B$, magnetic scalar potential $\varphi$ or magnetic vector potential $A$. In this study, we use magnetic scalar potential $\varphi$, because it has a set of simple boundary conditions.

We take a representative surface strip profile obtained from AFM characterization of a fine grain Nb surface. This surface is reasonably presumed isotropic at a scale of 100 µm since the typical fine grain Nb has grain size 20–50 µm. The boundaries of our model to describe the magnetic scalar potential, $\varphi$, near this surface are labeled with numbers in Figure 2. Boundaries 1 and 3 are a pair of periodic boundary conditions. We assign them Dirichlet boundary conditions where two arbitrary magnetic scalar amplitudes are given. These two values determine the applied parallel $H$ field far from the surface. This parallel $H$ field has a range from 80 mT to 210 mT in this simulation. Boundary conditions for boundaries 2 and 4 are the Neumann boundary conditions because they are treated as perfect electric walls. Boundary 4 is a surface characterized by AFM from a practical BCP-treated sample. In this analysis, we take $H_c$ = 190 mT. [14].

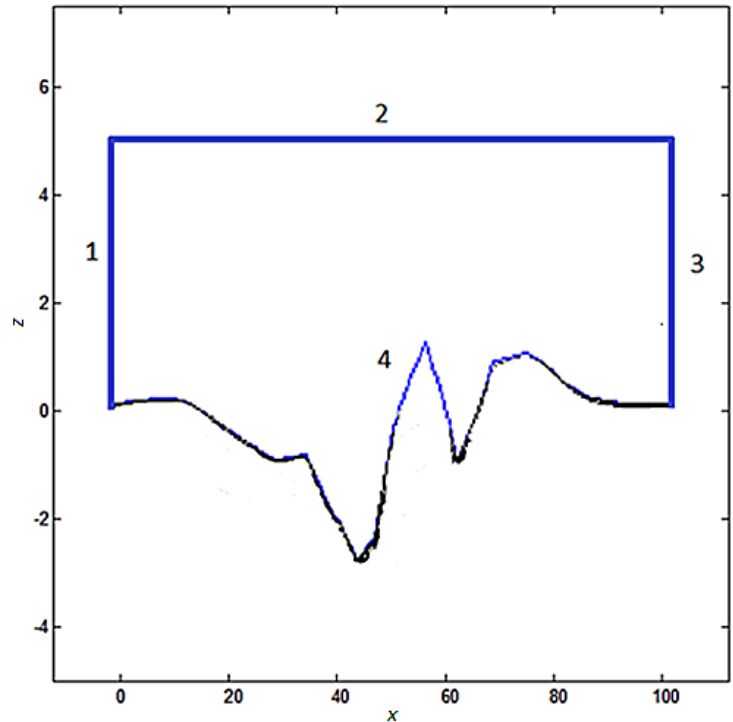

Figure 2: Configuration used for the simulation model calculation on a cross-section fragment of a BCP-treated sample surface. Area bounded in blue represents vacuum volume. Borderlines 1 and 3 determine the exciting magnetic field. Borderline 2 and 4 are perfect electric conductors

(PEC), while border 4 outlines a BCP treated surface profile, also PEC, acquired by AFM scanning. Unit: μm.

In Figure 3, the conformal mapping calculation yields the nominally horizontal solid lines as $E$ equipotential contours, while the vertical dotted lines are magnetic equipotential contours. The magnetic field at each point on the surface $H_{enhanced}(x)$ may be calculated as a function of horizontal position (x) by FEM. The solution obtained is the maximum amplitude of surface magnetic potential. The time dependent term must be added to represent the RF phase variation. The accuracy of the calculation is related to the surface characterization sampling and FEM resolution.

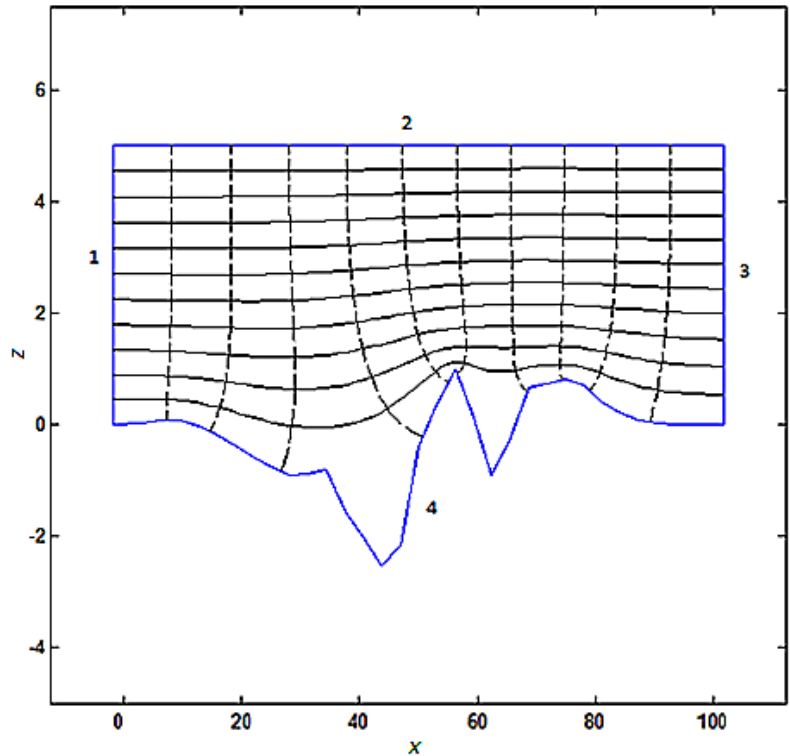

Figure 3: Electromagnetic equipotential contours by conformal mapping calculation. The vertical dashed lines are $E$ field lines, and horizontal lines are magnetic field lines. Unit: μm.

After FEM calculation, the magnetic field on the surface is computed by taking the derivative of the scalar potential along the surface. The LGMEF indexes along the horizontal x are plotted in Figure 4. The LGMEF factors are observed varying from 0.4 to 1.9. These amplitudes are attributed to the local surface topographic "sharpness." The LGMFE factor is greater than 1 on surface protrusions and smaller than 1 on valley areas.

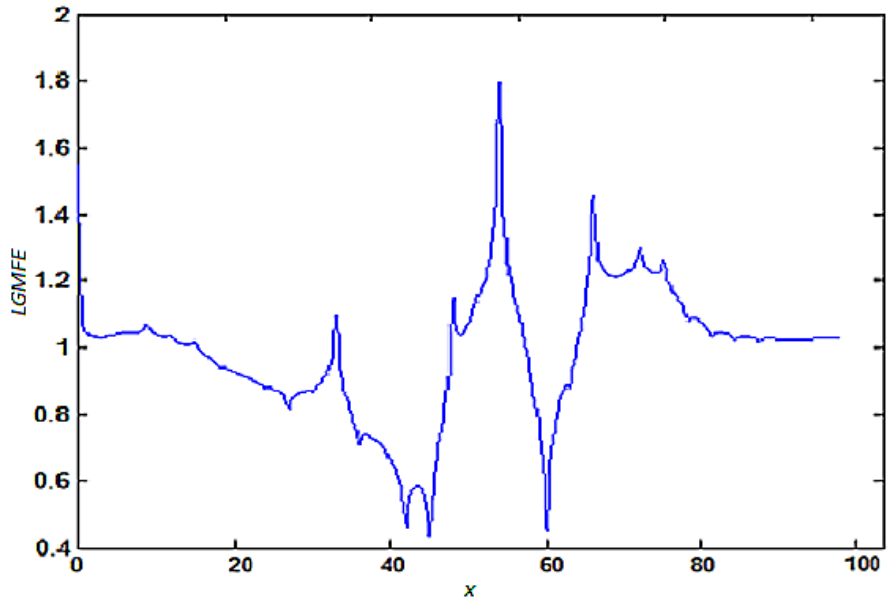

Figure 4: The LGMFE map is calculated from the profile in Figure 2. These indexes do not change with the applied field. Unit: µm.

When the applied magnetic field increases, the local $H$ field may begin to exceed $H_c$ at some surface areas where normal zones begin to nucleate. There then exists a normal and superconducting interface beneath the surface. In this study, we presume the superconducting to normal conducting transition would follows the change of the RF field magnitude. Thus, this interface moves inward and outward with RF phase. The location of this interface is determined by FEM at each snapshot. Mathematically, this situation is widely known as the Stefan moving boundary problem, and it simulates surface crystallization processes and other phase transition problems. [21] An additional borderline 5 is introduced on Figure 5. This outline 5 represents the interface, which we term the "normal conducting phase front." The tangential magnetic field value on this boundary is equal to the local $H_c$. Boundary 4 is subsequently ignored, because $H$ field decay within the normal zones between outlines 4 and 5, is negligible. The rest of the configuration in Figure 2 remains unchanged. Conditions on boundary 5 are expressed in equation 4. In the particular instance illustrated in Figure 5, the applied $H$ is greater than $H_c$ for clarity.

$$\begin{cases} H_\perp = 0 \\ H_\| = H_{critical} \end{cases} On\ Boundary\ 5 \quad \ldots\ldots\ldots\ldots\ldots\ldots\ldots\ldots\ldots\ldots\ldots\ldots\ (4)$$

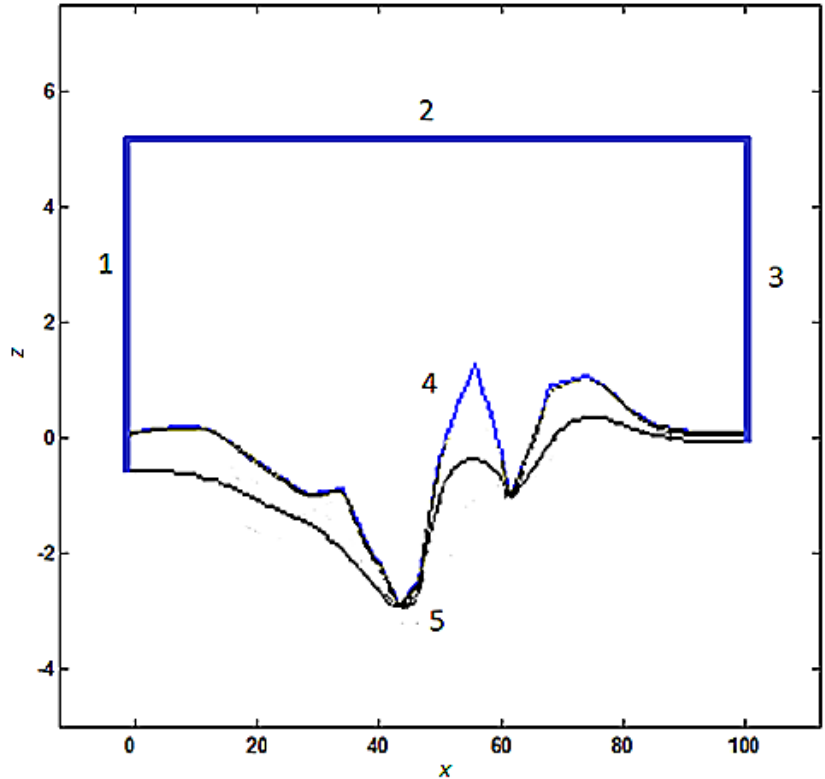

Figure 5: Configuration used for simulation model calculation on cross section fragment of the real BCP-treated sample surface. Area bounded in blue represents vacuum volume. Borderlines 1 and 3 determine the exciting magnetic field. Borderline 2 has PEC boundary condition.

Additional border 5 defines an interface of normal and superconducting materials and also is PEC. Unit: μm.

**Iteration method:**

Now let us discuss how to determine the location of this phase front. The basic algorithm is an iterative simulation until the known boundary conditions on the moving boundary are locally matched. The boundary conditions on this moving contour are listed in equation 4. Starting from the physical surface, one can calculate surface $H$ field at a given applied field. If any location on this changing boundary field has a local $H$ field greater than $H_c$, then the next step is to reduce this local surface height by a certain small amount. Continue calculating the field on this moving boundary until the local field on this boundary is equal to or less than $H_c$. Presumably, the $H$ field decay is negligible within the very shallow depth of the normal conducting zones. This is a reasonable assumption when the zone depth is a small fraction of the normal conducting skin depth. Localization of this phase front is thus calculated within certain error limitations. Accuracy can be improved at the expense of computing time. The iteration method is illustrated in the flowchart of Figure 6.

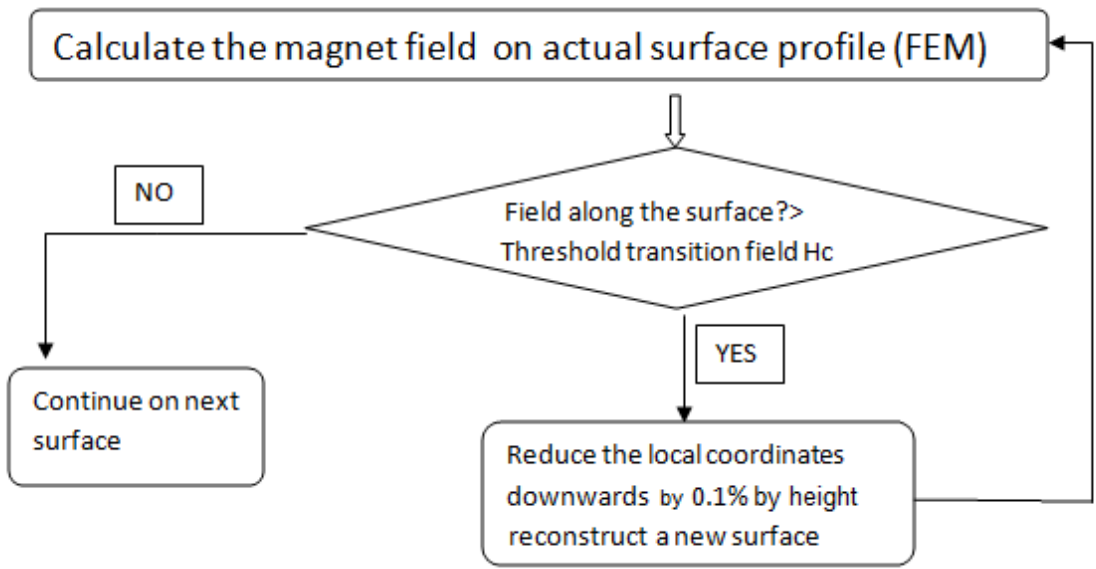

Figure 6: Flowchart to determine the interface between the normal conducting and superconducting conducting materials.

**Simulation results and comparison:**

Figure 7 provides the simulation results of the normal conducting phase front's deepest penetration when the amplitude of applied $H$ field ranges from 100 to 180 mT. The areas between the red and blue lines indicate the maximum normal zone volumes during each RF cycle.

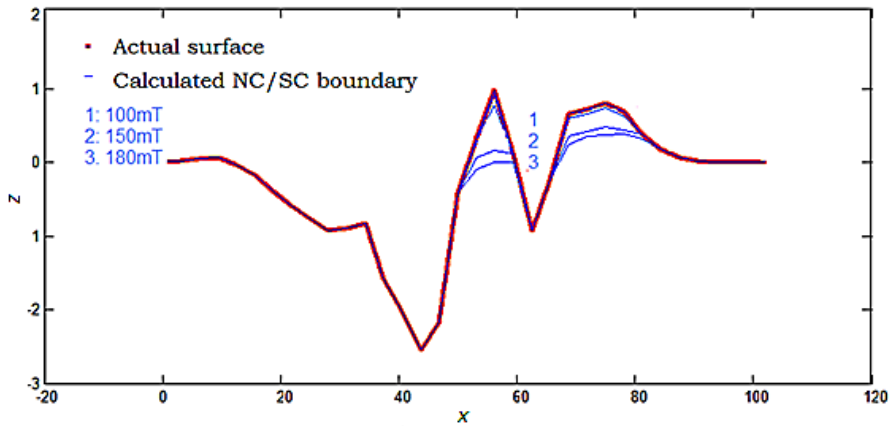

Figure 7: Normal conducting phase fronts as calculated from different excited fields. Unit: μm.

At low field, there is no normal zone because the local field is weaker than $H_c$. For example, if highest LGMFE index on a surface is 1.8 and $H_c$ is taken as 190 mT, the normal zone would be expected to nucleate when the applied $H$ field reaches 105 mT.

### 2.2 Thermal simulation and its correction iteration

**Heat equation:**

In this study, a thermal calculation uses the results from the electromagnetic simulation as input. This input includes the normal conducting/superconducting phase front location and $H$ field distribution. A goal for this thermal simulation is to generate a temperature map internal to the Nb from the RF surface to the external helium bath. The simulation estimates a temperature map in order to determine the material's phase, thermal conductivity, and dissipative losses in a self-consistent way.

After using the electromagnetic simulation results to obtain a temperature map, one can reassign the temperatures back to the material at each position to then determine the thermal conductivity. The change of thermal conductivity initiates the next round of temperature simulation. This iteration method may modify the normal-conducting phase front location results from the EM simulation, especially if the temperature of the normal conducting and superconducting interface is higher than $T_c$. In this thermal study, a second FEM computational code was developed to estimate the temperatures.

Thermal diffusion is governed by the partial differential equation:

$$\frac{\partial T}{\partial t} = div(\alpha(T) \cdot \nabla T) + \tilde{q} \quad \ldots \ldots \ldots \ldots \ldots \ldots \ldots \ldots \ldots \ldots \ldots \ldots \ldots \ldots (5)$$

where $T$ is temperature, $\tilde{q}$ is related to the internal heat source density, and $\alpha$ is the thermal diffusivity. Note that this diffusivity is a function of temperature.

Additionally, the internal heat source density can be further expressed as:

$$\tilde{q}(t,x,y,z) = \frac{\tilde{Q}(t,x,y,z)}{\rho C_P} \quad \text{and} \quad \alpha = \frac{\kappa}{\rho C_p} \quad \ldots \ldots \ldots \ldots \ldots \ldots \ldots \ldots \ldots \ldots \ldots (6)$$

where $\tilde{Q}$ is the heat generated at a given position and time, $\kappa$ is the thermal conductivity, $\rho$ is Nb density, $C_p$ is specific heat capacity, and $t$ is time. This heat is generated by RF loss on the surfaces.

For the static state solution, equation 5 reduces into:

$$div(\alpha(T) \cdot \nabla T) = \alpha(T)\nabla^2 T + (\nabla T)^2 \frac{\partial \alpha}{\partial T} = -\tilde{q} \quad \ldots\ldots\ldots\ldots\ldots\ldots\ldots\ldots\ldots\ldots \quad (7)$$

Note that the thermal conductivity is also temperature dependent.

With the first order solution, the thermal conductivity is a constant because the temperature difference on the surface is small. The equation 7 further reduces into a Poisson equation.

$$\alpha(T)\nabla^2 T = -\tilde{q} = -\frac{\tilde{Q}(t,x,y,z)}{\rho C_P} \quad \ldots\ldots\ldots\ldots\ldots\ldots\ldots\ldots\ldots\ldots \quad (8)$$

The right term $\tilde{q}$ in equation 8 is treated as a dynamic source, the RF power loss at a given field. The thermal diffusion time constant $t$ is determined by $\iota^2/\alpha$. The $\iota$ is characteristic size, which is 100 µm. The $\alpha$ is the thermal diffusivity, which is 5000 cm$^{-2}$sec$^{-1}$ at 4K.[22] Therefore, the thermal diffusion time constant is of order 10$^{-7}$ second. This means temperature change is slow compared with the RF field changes. The $\tilde{q}$ in equation 8 may then be an averaged thermal source, and the temperature map at an equilibrium state is calculated at a given field amplitude.

The thermal simulation setup is illustrated in Figure 8. The vertical simulated length is 3.3 mm, which is a typical cavity wall thickness. To confidently model the temperature map in a bounded area, the horizontal scale needs to be comparable to the cavity thickness. If the lateral length is set too small, the simulation leads to temperature calculation error because both side boundaries are heat isolation conditions. However, setting the lateral zone too large costs computation inefficiency. We take a lateral length of 6.6 mm in our simulation to simulate the thermal response of an isolated defect region under typical cavity cooling conditions. The geometry adaptive meshing technique computationally focuses attention on surface roughness features because the area ratio between roughness features (inserted) and the whole simulated area is small. [23]

Boundary conditions are illustrated in Figure 8. Borders 1 and 3 satisfy Neumann boundary conditions. Border 2 is the RF surface. The inset figure is an enlargement where the isolated surface feature for assessment with a lateral scale of 100 µm is located on the center of border 2. The grey area shows the heat source zone. The convection cooling boundary condition is applied at Border 4. Mathematically, it is a Robin or absorption boundary condition (ABC), and it can be expressed as below: [24]

$$\kappa \frac{\partial T}{\partial n} = h_{Kap}(T - T_{bath}) = \tilde{q} \quad \ldots\ldots\ldots\ldots\ldots\ldots\ldots\ldots\ldots\ldots\ldots\ldots\ldots (9)$$

where $\kappa$ is thermal conductivity, and $h_{Kap}$ is the Kapitza conductivity between helium and Nb. Both $h_{Kap}$ and $K$ are temperature dependent. These dependencies are given in equation 10. [25,26]

$$\kappa(T) = 0.7 e^{1.65T - 0.1T^2} \left(\frac{W}{K \cdot m}\right)$$

$$h_{Kap}(T, T_0) = 200 \cdot T^{4.65} \left[ 1 + 1.5\left(\frac{T - T_0}{T_0}\right) + \cdots \right. \\ \left. \cdots + \left(\frac{T - T_0}{T_0}\right)^2 + 0.25\left(\frac{T - T_0}{T_0}\right)^3 \right] \left(\frac{W}{K \cdot m^2}\right) \quad T - T_0 < 1.4K \quad \ldots\ldots(10)$$

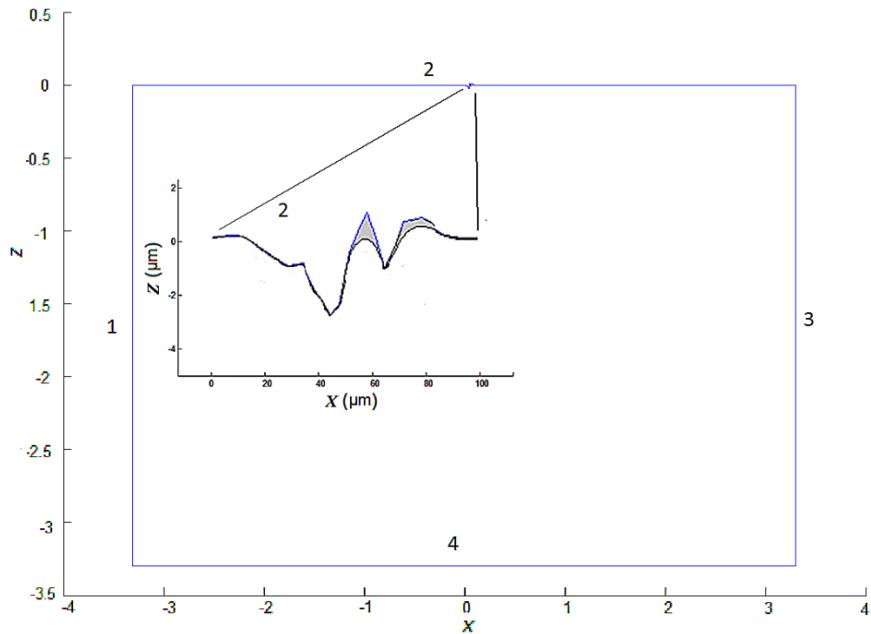

Figure 8: Typical temperature simulation area in 2D dimension: Borders are labeled in numbers. Borders 2 and 4 are RF surface and helium interfaces respectively, and Borders 1 and 3 have thermal isolation boundary conditions. Inserted figure: The surface roughness feature is highlighted. The grey area represents the internal heat source location. Unit: μm.

Next, we further consider the heat source term $\tilde{Q}$ in equation 8. Note that the commonly used surface area integration of equation 11 is applicable only if one presumes that the H field homogeneously penetrates the uniform surface within a skin depth. Our simulation is an unusual circumstance because the normal conducting dissipative layers are thinner than the normal conducting skin depth, unlike an assumption taken in [5]. This means equation 11 is not suitable for the loss calculation here.

$$\tilde{Q} = \int \tfrac{1}{2} \times R_{surface} \times H^2 \, dS \quad \ldots\ldots\ldots\ldots\ldots\ldots\ldots\ldots\ldots\ldots \quad (11)$$

Since this assumption is not valid in our simulation case, the RF dissipated from a small normal zone volume should be an integration based on the local electric field and electrical conductivity as in equation 12.

$$\tilde{Q} = \int \tfrac{1}{2} \times \sigma \times E^2 \, dV \quad \ldots\ldots \ldots\ldots\ldots\ldots\ldots\ldots\ldots\ldots (12)$$

where $\sigma$ is electric conductivity, $E$ is the volume electric field, and the integration $\tilde{Q}$ is the loss in the volume of the normal zone.

The electric field in the normal zone may be calculated from a quasi-static increasing $H$ field from Maxwell–Faraday law in equations 13:

$$\frac{\partial E_z}{\partial y} = \omega\mu H_x$$
$$\frac{\partial E_z}{\partial x} = -\omega\mu H_y \quad \ldots\ldots \ldots\ldots\ldots\ldots\ldots\ldots\ldots\ldots (13)$$

Note, in our model the electric field lies in a direction perpendicular to the plane of the paper and has integrated amplitude described by equation 14:

$$E_z(x,y) = \omega\mu \sin\omega t \left( \int_{y_0}^{y} -\frac{\partial \varphi}{\partial x} dy + \int_{x_0}^{x} \frac{\partial \varphi}{\partial y} dx \right) + E_0(x_0, y_0) \quad \ldots\ldots \ldots\ldots\ldots\ldots\ldots\ldots (14)$$

where electric field $E_0(x_0, y_0)$ is $E$ field on the normal conducting and superconducting interface. Its value is set to zero. Equations 13 and 14 suggest that RF power loss is proportional to $\omega^2$.

Numerically, RF power loss is calculated in the form of discrete power density on each element. This loss is the input for the thermal simulation. Compared to the RF loss in the normal zone, the RF loss from the superconducting zone is small and neglected at this stage. Thermal conductivity is updated locally after each iteration until a temperature map converges on each element. The algorithm is illustrated in the flowchart in Figure 9. With a converged temperature map, the resulting RF loss is expressed by an effective surface resistance.

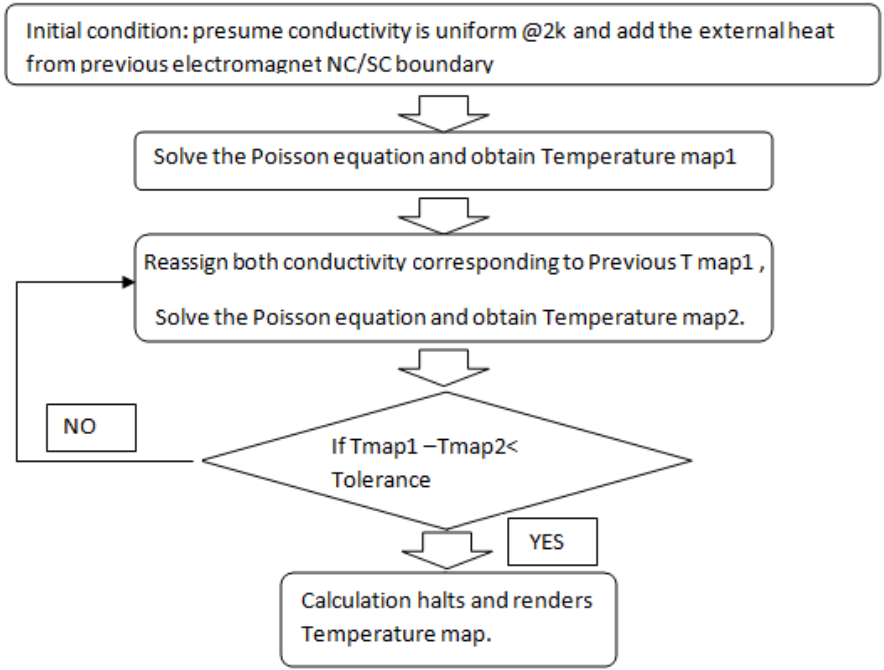

Figure 9: The flow chart of thermal equation simulation. It is used to calculate the temperature distribution.

**Simulation results and comparison:**

Using the results described in Figure 7, the calculation results of the consequent temperature map inside the cavity wall are demonstrated in Figure 10. The simulated setup configuration is from the model of Figure 8, and the results in Figure 10 are at two different applied $H$ field levels.

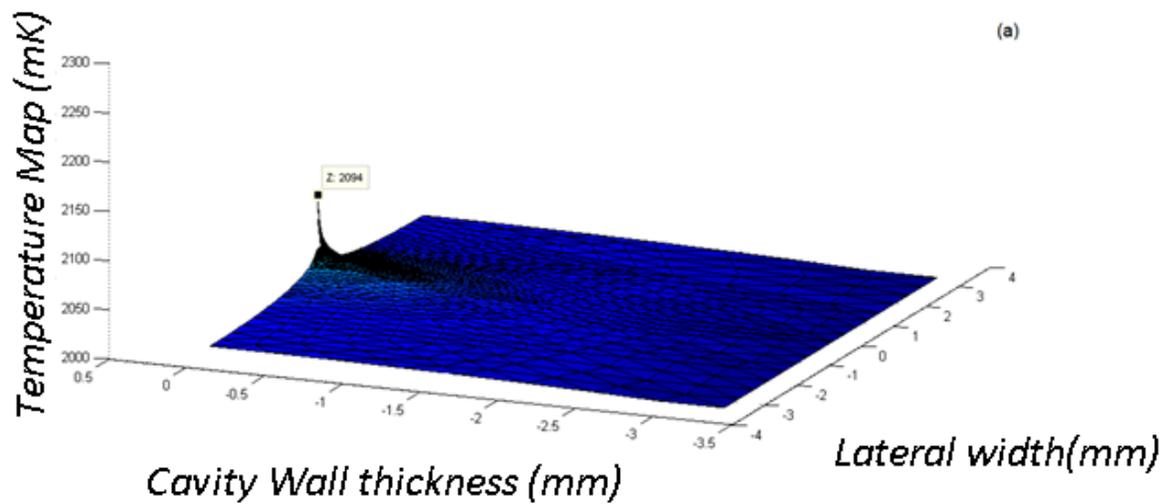

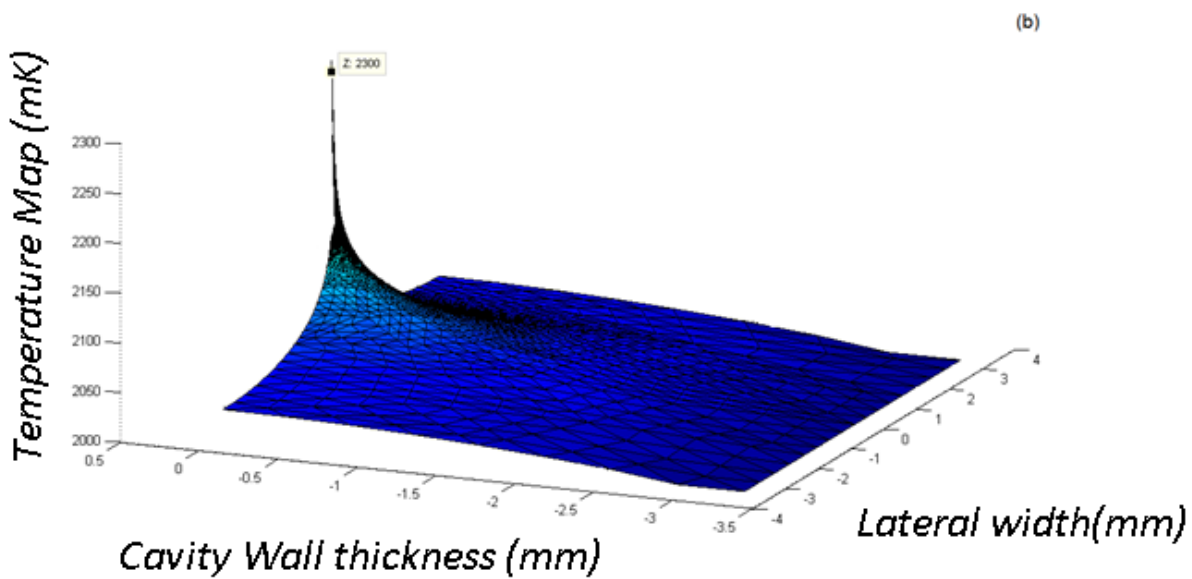

Figure 10: Temperature maps calculated for applied magnetic field of a) 100 mT and b) 120 mT with the isolated 100 µm rough strip from Figure 8. The maximum temperature reaches 2.094 K and 2.300 K, respectively. The helium bath condition is 2.00 K.

In Figure 10b, the radius of significantly heated zones on the surface can be as large as mm scale from a localized feature. With an exciting field of 100 mT, the highest temperature is calculated to be 94 mK higher than the helium bath temperature. At an exciting field of 120 mT, the highest temperature is around 300 mK higher than the helium bath temperature. These temperatures are far below the Nb transition temperature 9.2 K, suggesting that there is no significant thermal correction on the normal conducting zone size. It is thermally stabilized.

**2.3 Electromagnetic and thermal iteration simulation**

Superconductivity is bounded by three threshold critical parameters: current, magnetic field, and temperature. Temperature strongly influences the critical transition $H_c$ and further defines the normal conducting and superconducting interface location, which in turn determines the effective surface resistance. [1] Fortunately, $H_c$ varies little at low temperatures. Hence, this correction has a minor effect on RF loss estimate. $H_c$ (T) is typically corrected below in equation 15. [1]

$$H_c(T) = H_c(0K)[1-(\frac{T}{T_c})^2] \quad \ldots\ldots\ldots\ldots\ldots\ldots\ldots\ldots\ldots\ldots\ldots\ldots(15)$$

The location of the normal conducting phase front will be corrected numerically by the new temperature. Since the temperature rises at the sharp topographic features, local $H_c$ would decrease. Thus, a new electromagnetic and thermal configuration requires a recalculation. Therefore, we need to introduce a thermal feedback model including the $H_c(T)$ dependency and generate a higher level iteration that includes both simulations described in section 2.1 and 2.2. [22]. The flowchart of this big iteration is given in Figure 11.

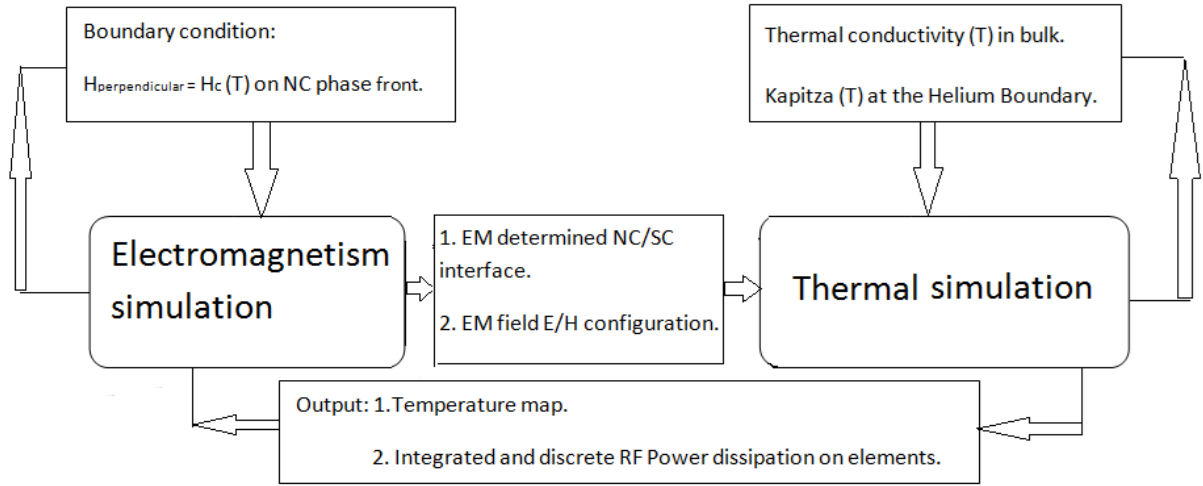

Figure 11: Flow chart of electromagnetic and temperature simulations. It is used to correct the size of the normal conducting zone and to estimate the RF loss. The flow chart shows a 'big' iteration with two 'small' iterations. Results of electromagnetic and thermal simulations are detailed in sections 2.1 and 2.2 in Figure 6 and Figure 9.

Similar to the representative BCP treated Nb surface in Figure 2, an electropolished (EP) fine grain Nb surface was characterized by a 100 µm AFM scan and is plotted in Figure 12. The same FEM calculation was conducted with the same boundary conditions described in Figure 2, only exchanging the boundary 4 with the representative EP surface profile. For this simulation, a geometry adaptive meshing was used to accommodate the fine surface features. The inset figure is an enlargement of meshing elements on the center of boundary 4 with an equal axis ratio.

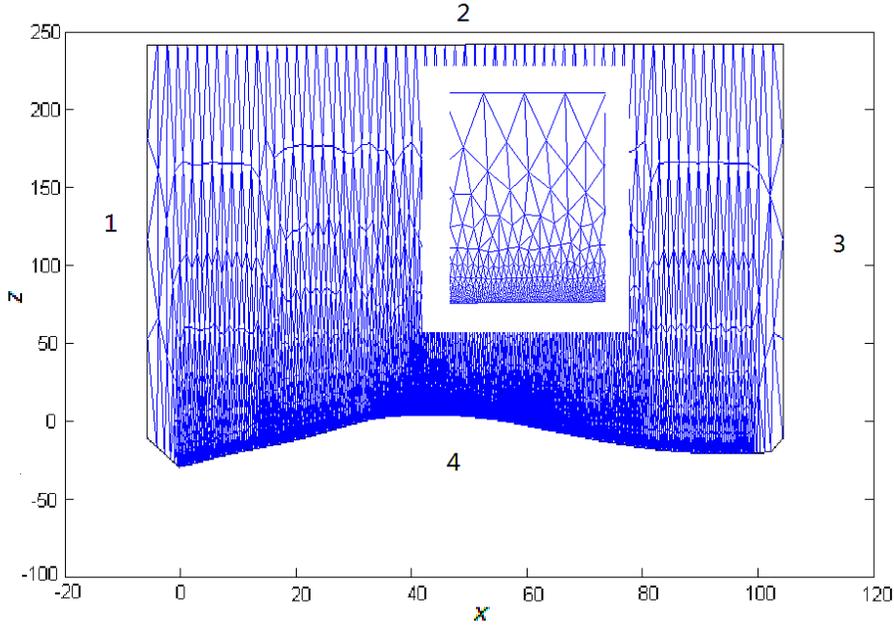

Figure 12: Electromagnetic calculation model for electropolished Nb surface. Note: the axis ratio is set as 3:1 to show in a distinguishable format. Insert figure is an enlargement where the axis ratio is 1:1. Unit: µm.

### 3. Application to characteristic etched and electropolished Nb surface topographies

The described integrated analysis above was applied to two 3D AFM profiles from BCP and EP treated fine grain Nb surfaces. Such representative surfaces can be replicated from cavities without undermining their performances. [27-29] The AFM scans used in this analysis are plotted in Figure 13. The AFM characterization area covers $100 \times 100$ µm. The effective raster strip width depends on the sampling rate, which is $512 \times 512$ in our case. Limited by computational capacity, we reduce the resolution to $32 \times 32$. As a result, strip columns, represented in Figure 2 and Figure 12, are taken to represent a width of 3.125 µm. The black line in Figure 13a marks such a typical strip. In this analysis, RF losses are then collected from the normal zones along 100 µm $\times$ 3.125 µm strips, and the resulting effective surface resistance increase from topographical field enhancements is calculated.

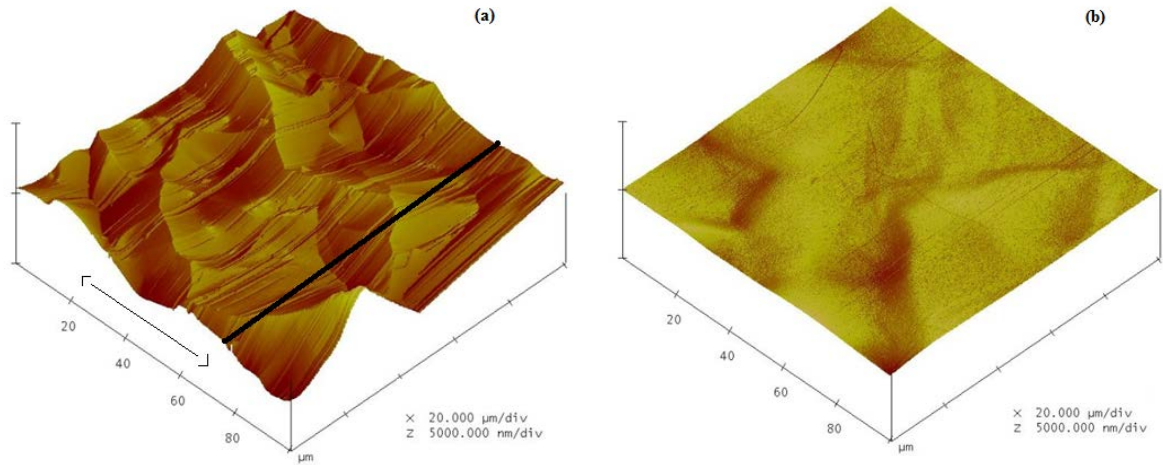

Figure 13. AFM images from a fine grain niobium sample with a) ~ 100 μm removal by BCP, b) after electropolished at 30°C to remove 48 μm. Horizontal scale is 20 μm per division and vertical scale is 5 μm per division. [3]

The RF losses on 32 such columns derived from the BCP surface profile were thus calculated as a function of applied $H$ fields. Figure 14 indicates the RF loss increasing with field due to the expansion of the small normal conducting zones. The losses from the individual strips (shown in blue) are calculated from equation 12. The averaged RF loss from these 32 strips is taken as representative of that due to the typical surface topography of a fine-grain Nb surface that has been etched by BCP. Note that all losses in the superconducting zones have been neglected here.

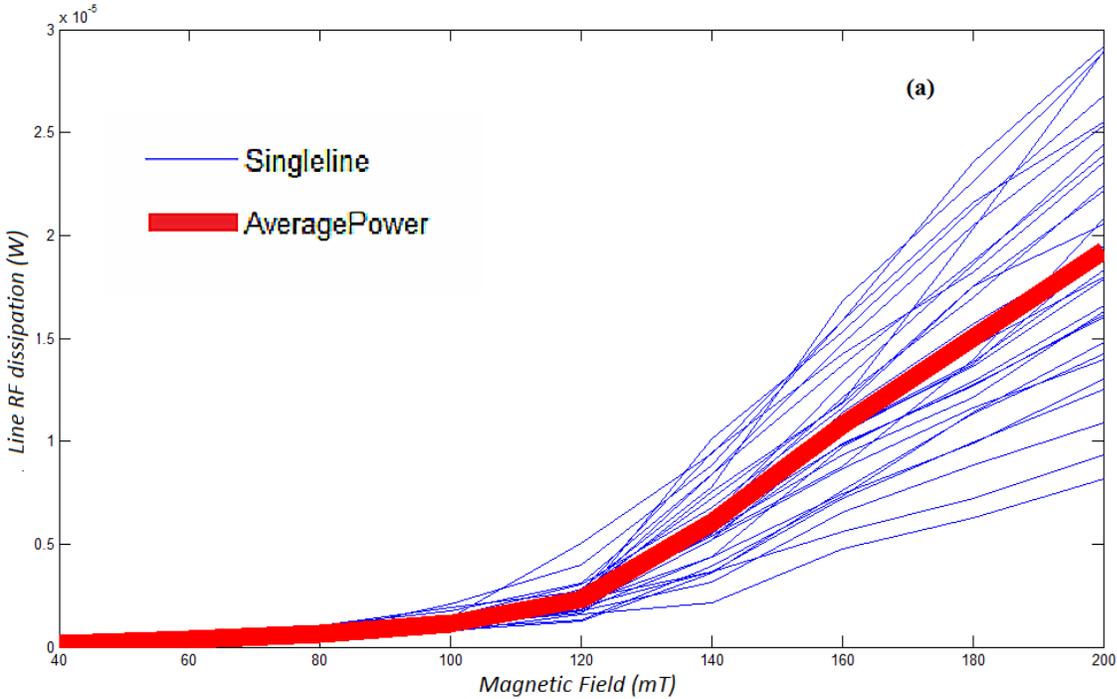

Figure 14. Calculated RF power dissipation on each representative 3.125 µm wide strip as a function of external applied magnetic *H* field for a 100 μm × 100 μm BCP treated fine grain Nb surface.  Blue lines are the RF loss from each of 32 strips and red line is the averaged RF loss

Figure 15 shows the average loss from an EP surface derived from the same analysis method as that shown in Figure 14 for the BCP'd surface.  Note the dramatic difference in calculated field-dependent losses from Nb surfaces etched by BCP and EP. These losses are collected from microscopic thermally stabilized normal conducting regions. Comparison of these two surfaces suggests that a significant density of small normal conducting zones is generated on BCP surfaces, while few normal zones are generated on EP treated surfaces. The additional heat generated on the EP-treated surface in this simulation is not significantly different from an ideally flat surface.

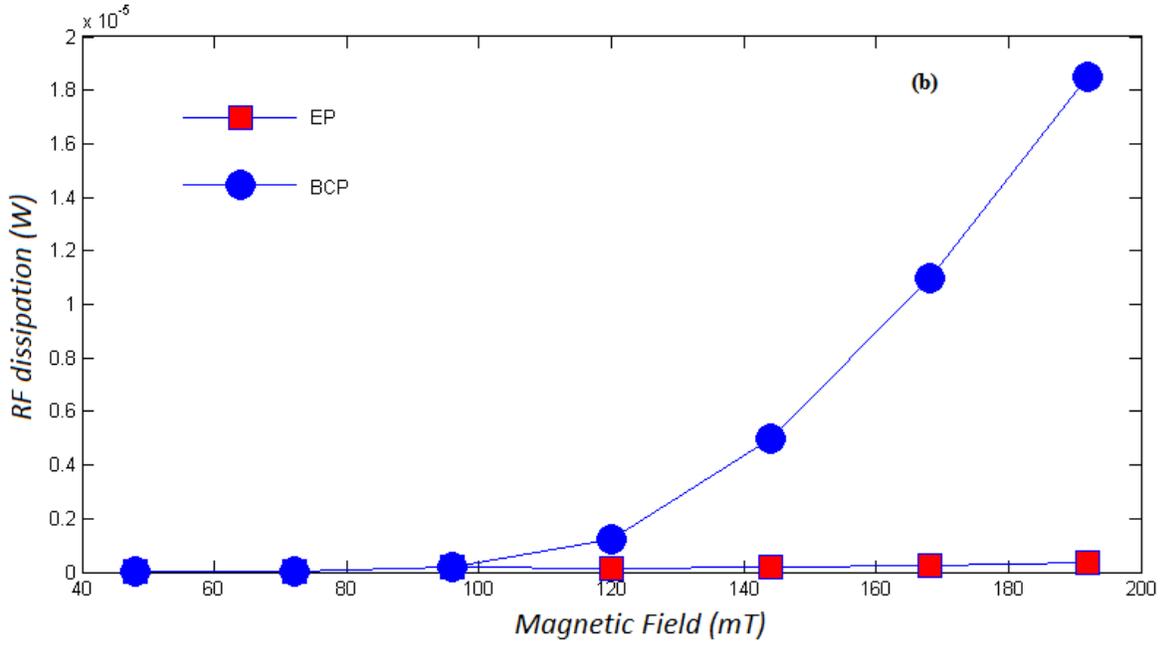

Figure 15. Average topography-induced power dissipation on 100 µm ×100 µm BCP and EP surfaces plotted as a function of peak applied *H* field. Superconducting state losses are ignored.

If the sum of RF losses on the 32 strips is represented by $\tilde{Q}$, the effective surface resistance is then :

$$R_{surface} = \frac{2\tilde{Q}}{H^2 \int ds} \quad \ldots\ldots\ldots\ldots\ldots\ldots\ldots\ldots\ldots\ldots\ldots\ldots\ldots (16)$$

This effective surface resistance is therefore proportionally related to the loss and density of microscopic field-induced normal zones. Now adding a small superconducting state resistance of ~13 nΩ, we obtain the resulting effective surface resistance as illustrated in Figure 16.

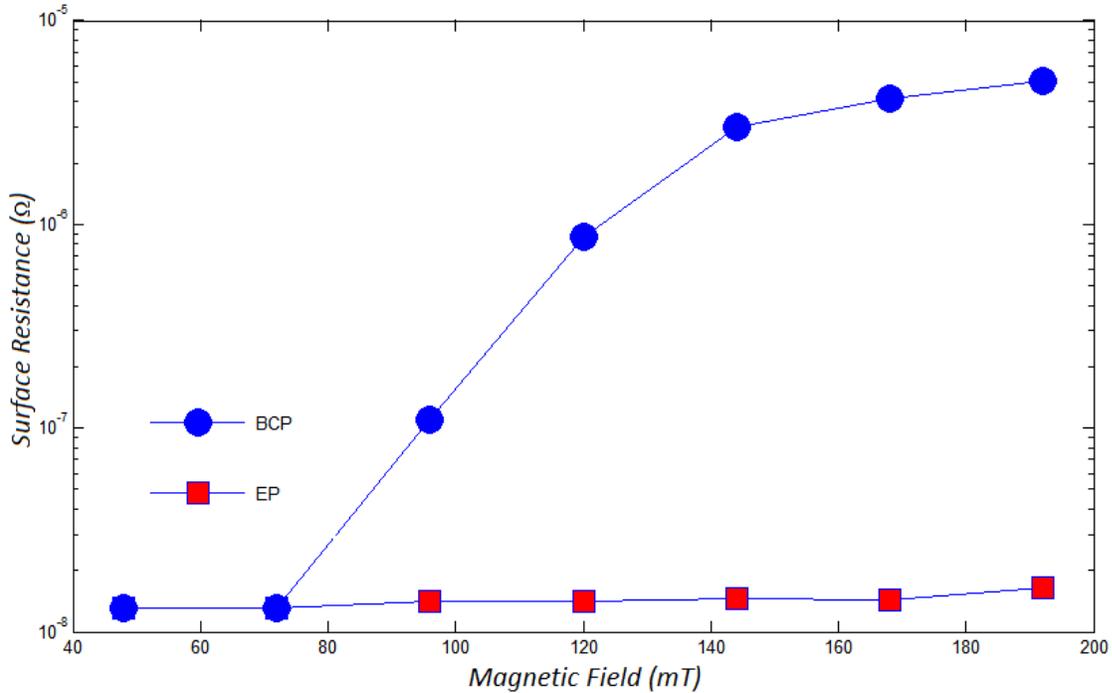

Figure 16: Simulated effective RF surface resistance with peak *H* field of BCP and EP treated fine grain Nb surfaces, including representative constant-temperature superconducting state losses.

**4.0 Discussion**

We now consider how such non-linear surface resistance would be reflected in the performance of a typical SRF accelerating cavity. Allowing that the local effective surface resistance has field dependency as described in Figure 16, we integrate the RF loss of a resonator cavity by equation 17.

$$P = \tfrac{1}{2} \times \int_{\substack{Cavity \\ Surface}} R(|H|) \times H(r,z)^2 \, dS(r,z) \; \ldots\ldots \ldots\ldots\ldots\ldots\ldots\ldots\ldots\ldots\ldots \; (17)$$

In a representative elliptical *β*=1 accelerating cavity, taking the approximation that the amplitude of surface *H* field is zero in the regions near irises and maximum along the equators, we can derive the effective surface resistance from the integration in equation 17. Then, the quality factor can be calculated from the simple expression of equation 18.

$$Q_{topo}(H) = \frac{\omega_0 U}{P_{heat}} = \frac{\frac{1}{2}\omega_0 \mu_0 \int_v H^2 dv}{\frac{1}{2} R_{Surface\ Resistance}(|H|) \times \int_s H^2 ds} \cong \frac{G}{R_{Surface\ Resistance}(H)} \quad \ldots\ldots\ldots\ldots (18).$$

The "Low Loss" cell geometry used in the CEBAF 12 GeV Upgrade 7-cell C100 cavity has been simulated in Superfish. [30] The normalized surface H field amplitude profile obtained is illustrated in Figure 17. In Figure 17, the blue curve is the cavity profile, the red curve is the surface $H$ field, and the dashed curve is our simplified surface field.

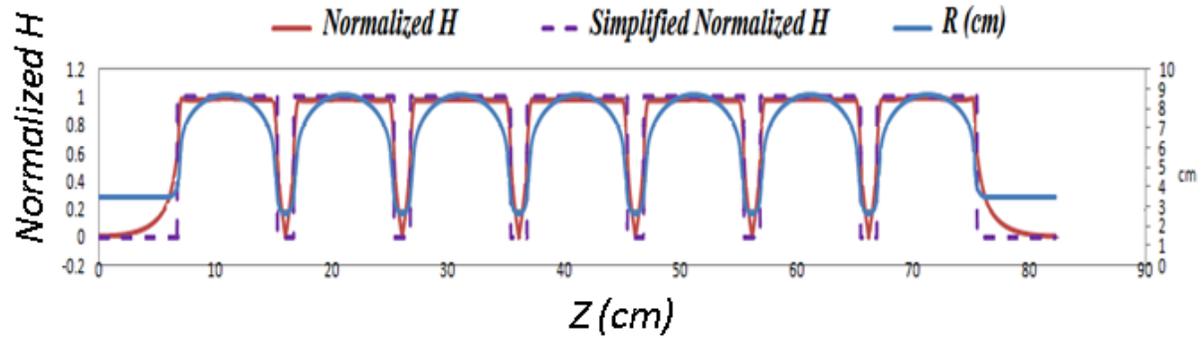

Figure 17: Surface magnetic field on CEBAF C100 7-cell cavity from Superfish.

The geometry factor for this structure in this accelerating mode is ~280 Ω and $B_{peak}/E_{acc}$ value is 3.74 mT/(MV/m). The quality factor of such a cavity with a correction for the interior surface topographic effect is given in equation 19.

$$Q_{calc} = \frac{G}{R_{surface\ resistance}(H)} = \frac{G}{R_{BCS} + R_{topo}(H)} \quad \ldots\ldots\ldots\ldots\ldots\ldots (19),$$

where the surface resistance comprises $R_{topo}(H)$ from Figure 16 and BCS resistance. The BCS surface resistance is presumed to have no field-dependency for 1st order simplicity. At 1.5 GHz, $R_{BCS}$ is commonly ~13 nΩ at 2 K, while the topographically-induced surface resistance is zero below some threshold field level. Consequently, the quality factor, $Q$, is dominated by BCS resistance at low fields. Figure 18 shows the quality factors, $Q_0$, as a function of surface $H$ field predicted by this analysis that would correspond to a "Low Loss" shaped cavity having Nb surface topography represented by the sampled two different surface treatments, BCP and EP. Note that thermal feedback on the surface resistance of the superconducting material has not been included, this would, of course, result in even further non-linear reduction of $Q_0$. As the normal conducting zones grow, some of the simplifying assumptions in our present analysis break down, the superconducting material losses become non-negligible, and the $Q$ decreases even faster than has been modeled here.

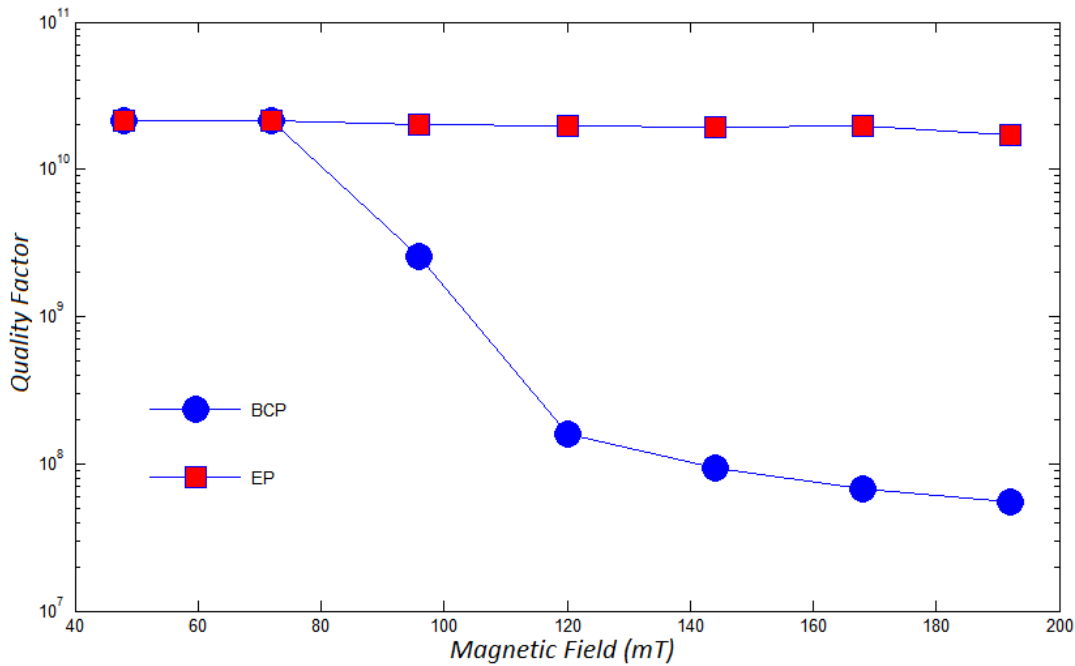

Figure 18: Comparison of the calculated effective cavity quality factor simulated for representative fine grain Nb BCP'd and EP'd surfaces at different peak *H* fields in a C100 geometry cavity. Thermal feedback effects are not included.

The model calculation results for the representative BCP-etched surface are in Figure 19 plotted together with 2 K performance test data for a Jefferson Lab upgrade prototype 7-cell cavity having this Low Loss geometry (LL002), both heavily BCP etched and subsequently electropolished.[31,32] Although one will certainly seek higher resolution from future model calculations, there is rough quantitative agreement between the calculation predictions and observed cavity performance in this case of a heavily etched cavity. This is consistent with the interpretation that topographical field enhancements are the cause of these enhanced non-linear losses on BCP-etched fine grain niobium, i.e. *Q*-drop from fine-scale roughness. Such field enhancements are absent from appropriately electropolished surfaces, so that this *Q*-drop mechanism is absent for EP-treated cavities.

Since the specific details of the surface structure of etched Nb surfaces (in contrast to electropolished surfaces) depend strongly on residual strains and defect densities, as well as the amount of etching removal from an otherwise smooth surface,[33,34] one should not be surprised to encounter significant variation of the topography-induced rf losses in different circumstances, though the phenomenon should be universal. One may, for example, interpret the small but significant systematic *Q*-drop reported at the high-field limit of the subset of XFEL cavities which received a final light BCP etch [35] as attributable to the low-amplitude sharpening of crystallographic edges creating widely dispersed local field enhancements.

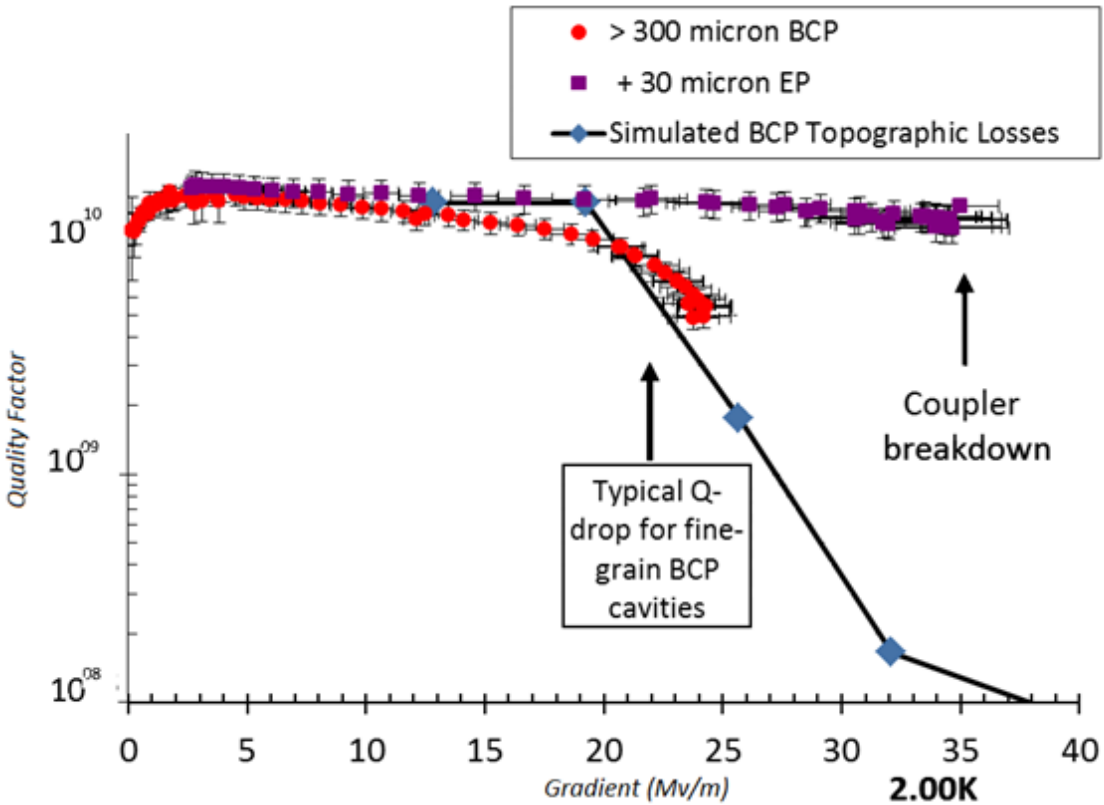

Figure 19: Comparison of model-predicted $Q_0$ having representative BCP surface topography with experimental data for CEBAF prototype cavity LL002 having heavily BCP etched and subsequent electropolished surfaces. No evidence of field emission loading was observed in either test.

## 5. Summary

Extending the analysis begun by [5], simplified electromagnetic and thermal simulations have been developed to analyze the microscopic scale geometric surface field enhancement and the normal/superconducting material interface when the local field exceeds $H_c$. The location of this interface phase front is a function of the exciting magnetic field and the specific topography. The thermally stable normal zone areas contribute significantly to the RF power loss. Each normal zone initiates its expansion based on the local geometric field enhancement factor. More accurate normal zone phase front modeling is obtained considering the critical field's temperature dependence. The volume of the normal zone on the surface expands nonlinearly with increasing external magnetic field. Such nonlinearity and the corresponding increase in dissipative RF power can be represented by an effective non-linear surface resistance. Additional superconducting RF loss is also induced as a result of thermal feedback. The total RF power loss contribution thus induced solely by topographic roughness may be calculated. Initial results of this analysis using representative topographic profile data from typical BCP etched and EP fine grain Nb surfaces yield a nonlinear loss character, and the *Q* dependence with field is quite similar to that typically observed with L-band SRF accelerating cavities with the corresponding surface treatments. This suggests that an adequate explanation for the field-dependent differences in BCP'd and EP'd fine-grain Nb cavities is now in hand. Opportunities to improve the precision of this model calculation have been identified, but physical insight into the phenomenon linking microscopic surface topography to high-field loss character of niobium SRF cavities for accelerators has been significantly enhanced.

**Acknowledgements:**

Authored by Jefferson Science Associates, LLC under U.S. DOE Contract No. DE-AC05-06OR23177. Chen Xu is grateful for discussions with Alexander Gurevich in Old Dominion University, Haipeng Wang, Hui Tian and Olga Trofimova at Jefferson Lab, John Delos at College of William and Mary and Toby Driscoll at University of Delaware.

**References:**

1. H. Padamsee, J. Knobloch, and T. Hays, *RF Superconductivity for Accelerators. 2nd Edition,*

(Wiley and Sons, New York, NY, USA, 2008).

2. C. E. Reece and G Ciovati, Rev. of Accel. Sci. and Technol. 5, 285 (2012).

3. C. Xu, H. Tian, C. E. Reece, and M. J. Kelley, Phys. Rev. ST - Accelerators and Beams. 14, 123501 (2011).

4. C. Xu, H Tian, C. E. Reece, and M. J. Kelley, Phys. Rev. ST - Accelerators and Beams. **15**, 043502 (2012).

5. J. Knobloch, R. L. Geng, M. Liepe, and H. Padamsee, Proceedings of 9th International workshop on RF Superconductivity, Santa Fe. NM. USA 1999, p. 77–91.

6. T. Kubo, Progr. Theor. Exp. Phys. 073G01 (2015).

7. E. Kako, S. Noguchi, M. Ono, K. Saito, T. Shishido, H. Safa, J. Knobloch, and L. Lilje, Proceedings of 9th International workshop on RF Superconductivity, Santa Fe. NM. USA. 1999, p. 179–186.

8. K. Saito, S. Noguchi, H. Inoue, M. Ono, T. Shishido, Y. Yamazaki, N. Ouchi, J. Kusano, M. Mizumoto, and M. Matsuoka, Proceedings of 8th Int. Workshop on RF Superconductivity, Abano Terme, Italy, 1997, p 534–539.

9. K. Saito, Proceedings of 2003 Particle Accelerator Conference, Portland, OR, USA, 2003, p 637-640.

10. D. Reschke, S. Aderhold, A. Gössel, J. Iversen, S. Karstensen, D. Kostin, G. Kreps, A. Matheisen, W.-D. Möller, F. Schlander, W. Singer, X. Singer, N. Steinhau Kühl, A.A. Sulimov, and K. Twarowski, Proceedings of 15th International Conference on RF Superconductivity, Chicago. IL. USA, 2011, p 490-494.

11. H. Tian and C. E. Reece, Proceedings of 15th International Conference on RF Superconductivity, Chicago, IL, USA, 2011, p 565-570.

12. C. E. Reece, A. C. Crawford, and R. L. Geng, Proceedings of 2009 Particle Accelerator Conferece, Vancouver, BC, Canada, 2009, p 2126–2128.

13. T.F. Stromberg and C. A. Swenson, Phys. Rev. Lett. 9, 370 (1962).

14. S. Posen, N. Valles, and M. Liepe, Phys. Rev. Lett. 115, 047001 (2015).

15. T. Kubo, Prog. Theor. Exp. Phys. 063G01 (2015).

16. S. Berry, C.Z. Antoine, and M. Desmons, Proceedings of European Particle Accelerator Conference, Lucerne, Switzerland, 2004, p.1000-1002.

17. V. Shemelin and H. Padamsee, Tesla Technology Collaboration Report No. 2008-07 Hamburg, German, 2008.

18. A. Dzyuba, A. Romanenko, and L. D. Cooley, Supercond. Sci. Technol. 23. 125011(2011).

19. S. Kim and I. E. Campisi, Phys. Rev. ST - Accelerators and Beams. 10. 032001 (2007)

20. J. D. Jackson, *Classical Electrodynamics. 3 edition,* (Wiley, New York, NY, USA, 1998)

21. S. I. Barry and J. Caunce, Appl. Math. Model. 32. 1. (2008)

22. J.E. Jensen, W.A. Tuttle, R.B. Stewart, H. Brechna, and A.G. Prodell, Brookhaven National Laboratory Report No. 10200 Volume II, 1980.

23. T. Plewa , T. Linde, and V. Gregory Weirs, *Adaptive Mesh Refinement, Theory and Application,* (Springer, Chicago, IL, USA, 2003).

24. K. Warnick, *Numerical Methods for Engineering: An Introduction Using MATLAB and Computational Electromagnetics,* (SciTech Publishing. Provo. UT. USA, 2011).

25. P. Bauer, N.Solyak. G.L.Ciovati, G. Eremeev, A.Gurevich, L.Lilje, and B.Visentin, Physica C. 441 51 (2006).

26. K. Mittag, Cryogenics. 13, 94 (1973).

27. C. Xu, C. Reece, and M. Kelley, Appl. Surf. Sci. 274, 15 (2013).

28. S. Berry, C.Z.Antoine, A.Aspart, J.P.Charrier, M.Desmons, and L.Margueritte, Proceedings of 11th workshop on RF Superconductivity, Lubeck, Germany, 2003, p591-593.

29. M. Ge, G. Wu, D. Burk, J. Ozelis, E. Harms, D. Sergatskov, D. Hicks, and L. D. Cooley, Supercond. Sci. Technol. 24, 035002 (2011).

30. J. Sekutowicz, G. Ciovati, P. Kneisel, G. Wu, A. Brinkmann, R. Parodi, W. Hartung, and S. Zheng, Proceedings of 2003 Particle Accelerator Conference, Portland, Oregon, USA, 2003. p 1395 – 1397.

31. C.E. Reece. E. F. Daly, S. Manning, R. Manus, S. Morgan, J. P. Ozelis, and L. Turlington, Proceedings of 2005 Particle Accelerator Conference, Knoxville, TN, USA, 2005 p 4081-4083.

32. C.E. Reece and H. Tian, Proceedings of 2010 Linac Conference, Tsukuba, Japan, 2010 p 779-781.

33. X. Zhao, G. Ciovati, C.E. Reece, and A.T. Wu, Proceedings of 2009 Particle Accelerator Conference, Vancouver, BC, Canada, 2009 p 2147-2149.

34. X. Zhao, G. Ciovati, C.E. Reece, and A.T. Wu, Proceedings of 2009 International Conference on RF Superconductivity, Berlin, Germany, 2009 p 446-448.

35. N. Walker, D. Reschke, J. Schaffran, and L. Steder, Proceedings of 2015 International Conference on RF Superconductivity, Whistler, BC, Canada, 2015.